\documentclass{article}
\usepackage[utf8]{inputenc}
\usepackage[normalem]{ulem}
\usepackage{graphicx} 
\usepackage{authblk}
\usepackage{xcolor}
\usepackage{tablefootnote}
\usepackage{cite}
\usepackage{amsmath,amssymb,amsfonts}
\usepackage{mathtools}
\usepackage{algorithmic}
\usepackage{textcomp}
\usepackage{relsize}

\usepackage[font=small,labelfont=bf]{caption}
\usepackage[square,numbers]{natbib}
\usepackage{hyperref}

\newcounter{bla}

\newcommand{\be}{\begin{equation}}

\newcommand{\ee}{\end{equation}}
\definecolor{codegreen}{rgb}{0,0.6,0}
\definecolor{codegray}{rgb}{0.5,0.5,0.5}
\definecolor{codepurple}{rgb}{0.58,0,0.82}
\definecolor{backcolour}{rgb}{0.95,0.95,0.92}

\usepackage{listings}
\definecolor{mGreen}{rgb}{0,0.6,0}
\definecolor{mGray}{rgb}{0.5,0.5,0.5}
\definecolor{mPurple}{rgb}{0.58,0,0.82}
\definecolor{backgroundColour}{rgb}{0.95,0.95,0.92}
\lstdefinestyle{CStyle}{
    backgroundcolor=\color{backgroundColour},   
    commentstyle=\color{mGreen},
    keywordstyle=\color{magenta},
    numberstyle=\tiny\color{mGray},
    stringstyle=\color{mPurple},
    basicstyle=\footnotesize,
    breakatwhitespace=false,         
    breaklines=true,                 
    captionpos=b,                    
    keepspaces=true,                 
    numbers=left,                    
    numbersep=5pt,                  
    showspaces=false,                
    showstringspaces=false,
    showtabs=false,                  
    tabsize=2,
    language=C
}

\title{Benchmarking and Parallelization of Electrostatic Particle-In-Cell for low-temperature Plasma Simulation by particle-thread Binding}
\author{Libn Varghese$^1$, Bhaskar Chaudhury$^1$, Miral Shah$^2$, Mainak
Bandyopadhyay$^{2,3}$}

\affil{$^1$Group in Computational Science and HPC, DA-IICT, Gandhinagar, India, 382007. \texttt{bhaskar\_chaudhury@dau.ac.in}}
\affil{$^2$Institute for Plasma Research, Gandhinagar, India, 382428}
\affil{$^3$Homi Bhabha National Institute, Anushaktinagar, Mumbai, India}

\date{ }

\begin{document}
\maketitle


\begin{abstract}
The Particle-In-Cell (PIC) method for plasma simulation tracks particle phase space information using particle and grid data structures. High computational costs in 2D and 3D device-scale PIC simulations necessitate parallelization, with the Charge Deposition (CD) subroutine often becoming a bottleneck due to frequent particle-grid interactions. Conventional methods mitigate dependencies by generating private grids for each core, but this approach faces scalability issues. We propose a novel approach based on a particle-thread binding strategy that requires only four private grids per node in distributed memory systems or four private grids in shared memory systems, enhancing CD scalability and performance while maintaining conventional data structures and requiring minimal changes to existing PIC codes. This method ensures complete accessibility of grid data structure for concurrent threads and avoids simultaneous access to particles within the same cell using additional functions and flags. Performance evaluations using a PIC benchmark for low-temperature partially magnetized \textbf{\textit{E $\times$ B}} discharge simulation on a shared memory as well as a distributed memory system (1000 cores) demonstrate the method's scalability, and additionally, we show the method has little hardware dependency.
\end{abstract}

%
\vspace{2pc}
%
%
%
%
\section{Introduction} \label{sec:intro}

Low-temperature plasmas (LTPs), typically partially ionized and containing a variety of ions, electrons and active molecules, are characterized by their unique thermal non-equilibrium and complex interactions, play a crucial role in a wide range of applications \cite{adamovich20222022,laroussi2017perspective,ovanesyan2019atomic,levchenko2020perspectives,speth2006overview}. Modelling and simulation of LTPs \cite{turner2017computer,taccogna2019latest,kim2005particle} are essential for enhancing scientific understanding and driving the development of novel applications~\cite{von2017foundations,Adamovich_2017, alves2018foundations} in diverse fields such as semiconductor processing \cite{oehrlein2018foundations}, medical research \cite{von2014clinical}, space exploration \cite{holste2020ion} and many more\cite{holste2020ion,samukawa20122012,Adamovich_2017}. Recently, there has been a growing interest in investigation of LTP-based \textit{\textbf{E $\times$ B}} devices \cite{kaganovich2020physics} such as hall thrusters \cite{adam2008physics,boeuf2017tutorial,levchenko2020perspectives,perales2022hybrid} and negative-ion sources \cite{shah2023investigation,kolev2012physics,schiesko2012magnetic,garrigues2016appropriate} which typically operate at relatively low gas pressures. The external magnetic field in such devices is usually applied perpendicular to the plasma flow, inducing a drift in the \textit{\textbf{E $\times$ B}} direction that leads to various instabilities \cite{smolyakov2016fluid,hara2019overview}. The presence of an external magnetic field, various types of collisions, and plasma inhomogeneities result in highly complex, multiscale physics that require kinetic simulations for thorough analysis. Particle-In-Cell (PIC) \cite{donko2021edupic,arber2015contemporary,decyk2014particle,burau2010picongpu,buneman1959dissipation,6808530,kim2005particle,derouillat2018smilei} method is a well-established and accurate computational approach used for large-scale kinetic simulations of different kinds of plasmas \cite{birdsall2004plasma}. 

The Electrostatic Particle-In-Cell (ES-PIC) simulations \cite{SUN201635,garrigues2016appropriate,tskhakaya2007particle}, are used for studying LTPs where the time derivative of the magnetic field is not relevant. Many different implementations of ES-PIC are available; however, the basic flow of the implementation remains the same. There are two primary data structures for a standard ES-PIC code \cite{chaudhury2019hybrid}, firstly, \textit{Particle} data-structure, which stores particle information on the  Lagrangian grid, which has the $x, y, z$ for location, $v_x, v_y, v_z$ for velocity and particle type (electron, ion etc.) and secondly, \textit{Grid} data-structure, which contains the electric field, magnetic field, potential and charge density calculated on the Euler Grid, which mediates the collective interaction among the particles. ES-PIC has three critical modules: 1) Particle-To-Grid linear interpolation of charge density from the particle to the grid (\textit{Charge Deposition - CD}), 2) Calculation of potential and electric field, 3) Grid-to-Particle interpolation to evolve particle location and velocity based on the fields on the grid (\textit{Mover}). The main loop keeps iterating in the order 1-2-3; a detailed implementation of ES-PIC can be found here \cite{BChaudhury2017}. Runtimes can vary significantly based on implementation and hardware \cite{adams2007performance,charoy20192d}. Due to the immense computational load, most 2D and 3D ES-PIC require innovative parallelization techniques to take full advantage of high-performance computing (HPC) systems \cite{miller2021dynamic,deluzet2023efficient,tang2016extreme,wright2024developing,gruber2023llama,incardona2019openfpm,burau2010picongpu,Hur_2019,taccogna2023plasma,jambunathan2018chaos,juhasz2021efficient}. Recent benchmarking efforts by seven groups worldwide have focused on the development, validation, and acceleration of \textit{\textbf{E $\times$ B}} based LTP simulations, demonstrating that these simulations require tens of thousands of CPU hours. \cite{charoy20192d}.

It is known that \textit{Mover} has good scalability \cite{LIEWER1989302}. However, the \textit{CD} module has been the bottleneck \cite{vincenti2017efficient,steiniger2023ez,burau2010picongpu,claustre2013particle,decyk2011adaptable,abreu2010pic} for most parallel PIC codes because dependency issues prevent effective parallelization.  In the conventional method for parallelizing \textit{CD}, private grids associated with charge-density are generated for each processing core, as described in \cite{chaudhury2019hybrid}. However, this approach does not scale well with an increasing number of cores. A notable contribution in this area was reported by Stanchev et al. \cite{Stantchev2008}. Subsequently, over the past decade, limited work has been done to improve its performance further. In this context, we introduce a particle-thread binding (PTB) approach to enhance the scalability and performance of \textit{CD} while preserving the fundamental data structure associated with the conventional ES-PIC. A sorted particle data structure is required when using only four private grids, as all particles in the same cell must be assigned to the same thread. In the proposed approach, we employ a PTB mechanism to avoid sorting the particle data structure at every iteration while ensuring that only four private grids are utilized. While the proposed method is primarily designed for ES-PIC, it can also be applied to other systems that require repeated conversion of mesh-free scattered points to structured grids \cite{hockney2021computer,madduri2012optimization,peskin1989three,sorensen2008accelerating}. To test the performance and scalability of the proposed method, we conducted experiments on shared memory and distributed memory systems. Our experiments show encouraging results and excellent scalability on an HPC platform with 1000 cores.

This paper is structured as follows: Section \ref{Physics} provides an overview of the underlying physics of the problem, the numerical implementation and the fundamental flow of PIC. We validate the accuracy and performance of our in-house developed 2D-3v parallel PIC code by presenting the results of a 2D benchmark problem \cite{charoy20192d}. Section \ref{Methodology} provides a detailed explanation of the parallelization strategies used, focusing on the challenges associated with parallelising the charge deposition module, which traditionally hinders the scalability of PIC. A novel PTB method is introduced to enhance the scalability of this module and improve the overall performance of the PIC simulations. Finally, Section \ref{Results} presents the performance improvements observed from numerical experiments across various problem sizes and on two different architectures, followed by an analysis of the results.

\section{Physical and Computational model and implementation detail }\label{Physics}

\subsection{Electrostatic PIC for Low-Temperature Plasma simulation}
\begin{figure}
    \centering
\includegraphics[width=1.0\textwidth]{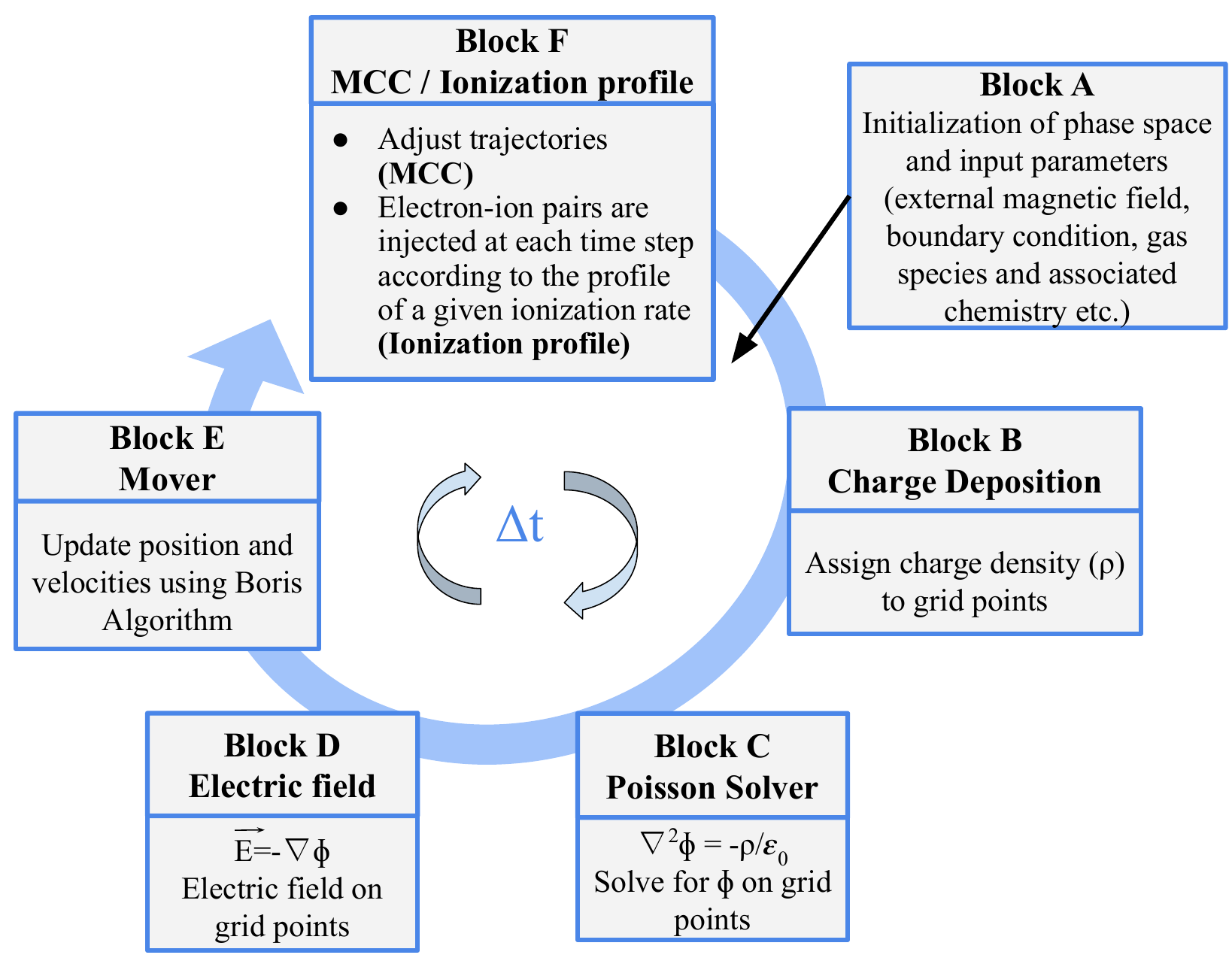}
    \caption{Self-Consistent Particle-in-Cell Simulations time loop with Poisson Solver, Charge Deposition, and Monte Carlo Collision/ Analytical Ionization profile}
    \label{fig:flow chart}
\end{figure}

The flowchart illustrating the self-consistent ES-PIC \cite{garrigues2016appropriate,birdsall2004plasma,derouillat2018smilei,tskhakaya2007particle} method is depicted in Fig. \ref{fig:flow chart}. In a ES-PIC simulation, the spatial and temporal evolution of representative particles known as superparticles and fields are tracked \cite{kim2005particle,hara2023effects}. The phase space information of these superparticles is tracked on a continuous Lagrangian grid, while field information is tracked on the discretized Euler grid across the simulation domain. The interaction of these superparticles determines the macroscopic nonlinear dynamics of plasma \cite{donko2011particle}.  Typically, the key computationally demanding modules of ES-PIC are \textit{Charge Deposition}  (Block B), \textit{Mover} (Block E), and \textit{Monte Carlo Collision (MCC)/Ionization} for plasma chemistry  (Block F). Particle trajectories are evolved by solving Newton's equations, typically expressed as a system of first-order ordinary differential equations. Determining charge density $(\rho)$   involves interpolating from the Lagrangian grid onto an Euler grid via particle-to-grid interpolation, done in \textit{CD}. Subsequently, numerically solving Poisson’s equation (Block C), represented by Eq. \ref{poisson equation}, the potential $(\phi)$ on the grid is computed, which is then used to compute the electric field $(\vec{E})$(Block D) using Eq. \ref{grad e}.
\begin{equation}
    \nabla^{2}\phi = -\frac{\rho}{\epsilon_0}
    \label{poisson equation}
\end{equation}
\begin{equation}
    \vec{E} = -\nabla\phi
    \label{grad e}
\end{equation}

The equation governing the position $(\vec{r})$ and velocity $(\vec{v})$ of particles is given below:
\begin{equation}
    m\frac{d\vec{v}}{dt} = \vec{F}
    \label{velocity update}
\end{equation}
\begin{equation}
    \frac{d\vec{r}}{dt} = \vec{v}
    \label{position update}
\end{equation}

Force, $(\vec{F})$, contains two distinct components representing the influence of electric $(\vec{E})$ and magnetic $(\vec{B})$  fields $\vec{F}=q(\vec{E}+\vec{v}\times \vec{B})$. The charge and mass of a particle are given by $q$ and $m$, respectively. Computing this force involves determining the fields at the specific position of each particle. This computational step necessitates the interpolation of fields from an Euler grid to the Lagrangian grid within the designated time step, via grid-to-particle interpolation done in \textit{Mover}. The grid size must be comparable to the Debye length ($\leq \lambda_{D}$) and time-step $\Delta t \leq \frac{0.2}{\omega_{p}}$ for PIC stability, where plasma frequency $\omega_p = \sqrt{\frac{n_e e^2}{m_e \epsilon_0}}$ and the electron Debye length $\lambda_d = \sqrt{\frac{\epsilon_0 K_B T_e}{n_e e^2}}$. $n_e$ being the electron density, $e$ the electron charge, $m_e$ the electron mass, $T_e$ the electron temperature and $\epsilon_0$ the vacuum permittivity. This makes device scale PIC simulations a computationally challenging problem. Implementation details of PIC can be found here \cite{chaudhury2019hybrid}.  

\subsection{Data structure and implementation of our in-house 2D-3v PIC code} 
\begin{figure}
    \centering
\includegraphics[width=1.0\textwidth]{ 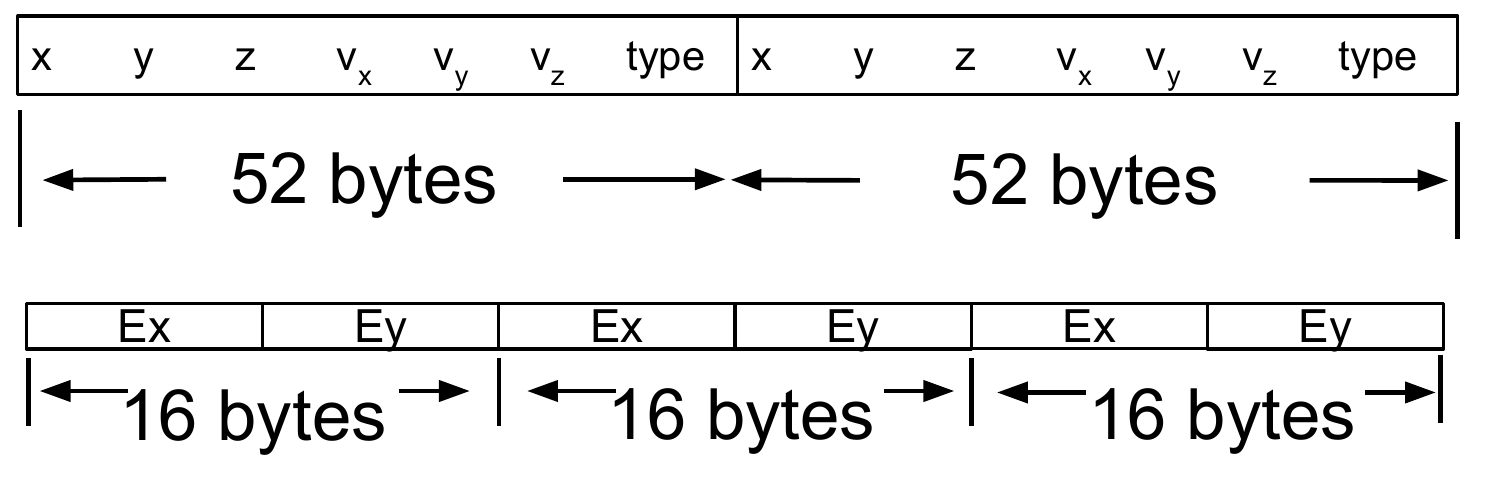}
    \caption{Data structure representing the storage layout of the \textit{Particle} (Lagrangian) and \textit{electricField} (Eulerian) grids in memory.}
    \label{fig:Data structure}
\end{figure}

The phase space information of each particle, including position $(\vec{r})$ and velocity vectors $(\vec{v})$, is stored as an array-of-structure (AOS)  named \textit{Particle}, along with their respective species identifier integer (electron or ion) as depicted in Fig. \ref{fig:Data structure}. Quantities such as charge density, potential, electric, and magnetic fields are discretized over the 2D grid. For an electric field, each grid point features two components, one along the x and the other along the y direction. Consequently, electric field information is stored in a structure dedicated to a particular grid point, and an AOS named \textit{electricField} maintains the electric field data for the entire grid. Currently, the magnetic field in our simulations varies solely in the x-direction, necessitating only an array of values to track it across all grid points. All quantities, whether particle or grid-related, are stored as double-precision floating-point values during simulation. Notably, despite grid quantities being distributed over the points of a 2D grid, we employ a 1D array to store them \cite{chaudhury2019hybrid}. Poisson’s equation is solved using the direct PARDISO solver included in the MKL library of INTEL\cite{schenk2004solving} to get the potential on the grid. The Eq. \ref{velocity update} and Eq. \ref{position update} are solved using a numerical integration scheme, the time-centred leapfrog Boris method \cite{boris1970relativistic}. The CPU version of the PIC code is written entirely in C and employs two levels of parallelism. The first level of parallelism is node-level, implemented using the Message Passing Interface (MPI). The second level is thread-level parallelism, achieved using OpenMP. MPI-based communication becomes effective when multiple nodes are used, and communication is needed among them, particularly in distributed memory systems. This allows the simulation to scale efficiently across multiple nodes. OpenMP is utilized for shared memory systems, enabling parallel execution within a single node by distributing tasks among multiple cores. The codes are compiled using the Intel MPI ICC compiler with appropriate flags, which supports both MPI and OpenMP. This setup allows for flexibility, as the number of nodes and cores can be specified at runtime, facilitating easy simulation scaling to match the available hardware resources. 

\subsection{High-performance computing platform}
 To test our algorithm, we used the ANTYA HPC facility with thousands of cores based on the philosophy of distributed memory, at the Institute for Plasma Research, India. This HPC features 40 cores on each node, distributed across two sockets with 20 cores each. Each socket has an Intel Xeon Gold processor with 32 KB L1, 1 MB L2 cache per core, and 27.5 MB shared L3 cache. Each node is equipped with 376 GB of total memory with Non-Uniform Memory Access  (NUMA). 
Each node is interconnected using a high-speed interconnect with a peak throughput of 100 Gbps. The combined capabilities of these nodes give ANTYA a peak performance throughput of 1 petaflop, making it an ideal platform for testing and benchmarking advanced PIC codes that require substantial computational power and parallel processing capabilities.

\subsection{Benchmarking of In-house developed 2D-3v massively parallel PIC code}
\label{benchmarking}

\begin{figure}
    \centering

\includegraphics[width=1.0\textwidth]{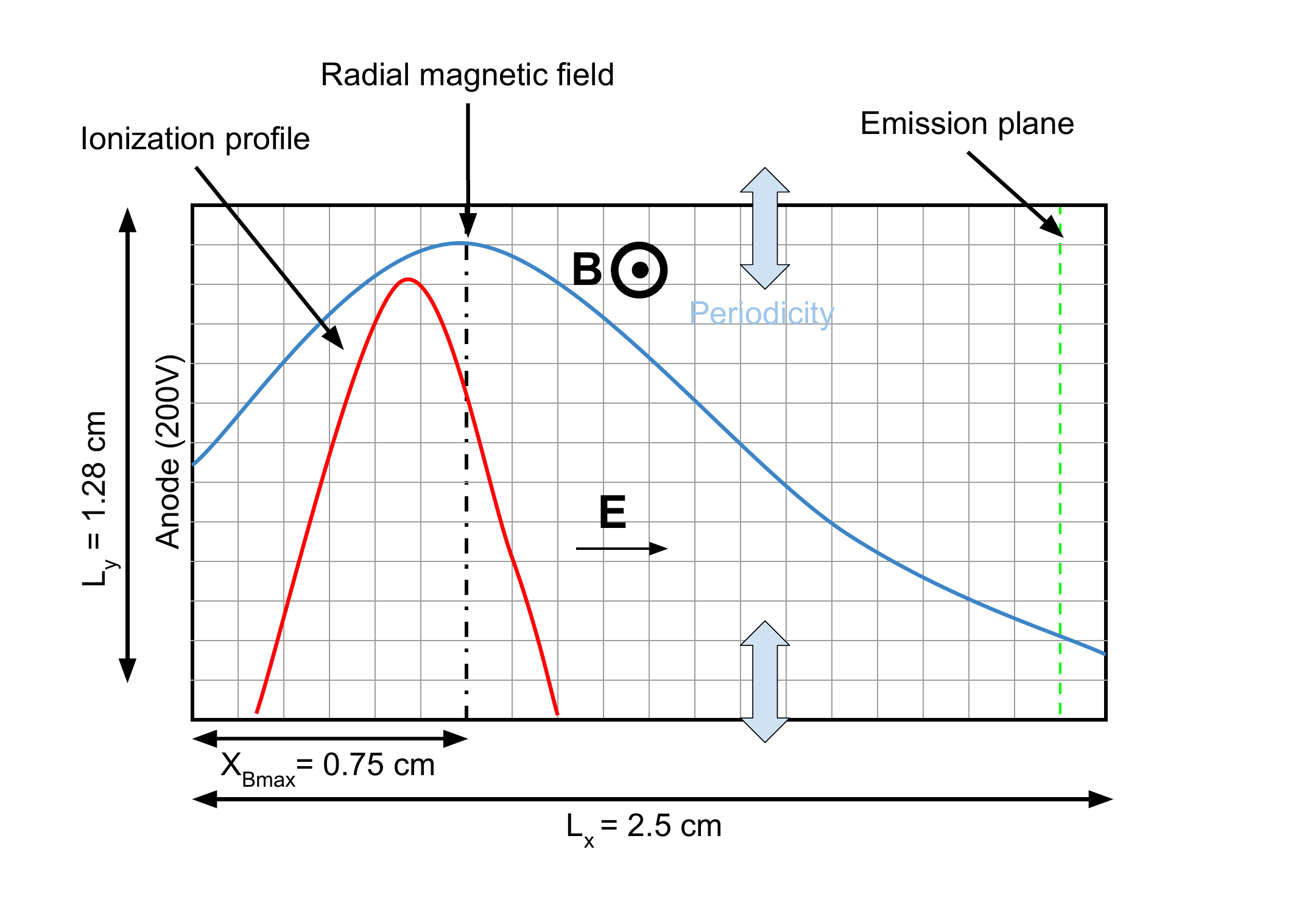}
    \caption{The left boundary of the domain is set at a constant potential of 200V, the right boundary set at 0v. Axial length is $L_x=2.5$ cm, with the radial magnetic field peaking at $x=0.75$ cm, which is directed into the simulation domain in the z direction. $L_y=1.28$ cm}
    \label{fig:Simulation domain}
\end{figure}

Ensuring accuracy amidst the complexity of numerical models necessitates rigorous validation and verification \cite{turner2013simulation,villafana20212d,alves2023foundations}. To validate the accuracy and performance of our PIC code, we undertake the benchmark problem that investigates the azimuthal \textit{\textbf{E $\times$ B}} electron drift instability and the associated axial electron transport \cite{charoy20192d,boeuf2018b,villafana20233d,charoy2020comparison}. Here, no collisions are considered; instead, a prescribed ionisation source term sustains the plasma discharge, which leads to a steady state within a brief computational timeframe, specifically within $10\mu s$. The axial profile of the radial magnetic field is shown in Fig. \ref{fig:Simulation domain}. The applied magnetic field is perpendicular to the plasma flow.
This test case is designed to study the physics of a Hall Thruster and serve as a general benchmark\cite{reza2023concept} for all \textit{\textbf{E $\times$ B}} discharge codes.
A 2D structured Cartesian mesh representing the axial (x) and azimuthal (y) directions of this problem is illustrated in Fig. \ref{fig:Simulation domain}. Table. \ref{tab:parameter} enlists all the input parameters for the benchmark problem, for further details refer \cite{charoy20192d}. Seven independently developed codes from around the world were used to simulate the benchmark problem \cite{charoy20192d}, highlighting distinct features of these codes and evaluating the convergence of results. On average, these codes took 60,000 core hours for complete simulation, highlighting the necessity for massive parallelization strategies and the development of new algorithms to reduce computational costs \cite{derouillat2018smilei}. 

\begin{table}[htbp]
  \centering
  
  \begin{tabular}{|c|c|c|}
    \textbf{Physical parameter} & \textbf{symbol} & \textbf{value}\\
   
   Initial plasma density & $n_e$ & $5\times 10^{16}m^{-3}$\\
   Discharge voltage & $U_0$ & $200$V\\
   Electron initial temperature & $T_e$ & $10$eV\\
   Ion initial temperature & $T_i$ & $0.5$eV\\
   Axial length & $L_x$ & $2.5$cm\\
    Azimuthal length & $L_y$ & $1.28$cm\\
    Final time & $T_{final}$ & $20\times 10^{-6} s$\\
   \hspace{5pt}
   \textbf{Computational parameter} & \textbf{symbol} & \textbf{value}\\
   Initial particles per cell & $ppc$ & $75$\\
   Time step & $\Delta t$ & $5\times 10^{-12} s$ \\
    Cell size & $\Delta x$, $\Delta y$ & $5\times 10^{-5} m$\\ 
    Total time steps & $steps$ & $4\times 10^{6} s$\\
     Cells in axial direction & $N_x$ & $500$\\
   Cells in azimuthal direction & $N_y$ & $256$\\
   
  \end{tabular}
  \caption{Important parameters for benchmark low temperature plasma simulation}
  \label{tab:parameter}
\end{table} 



\begin{figure}
    \centering
\includegraphics[width=1.0\textwidth]{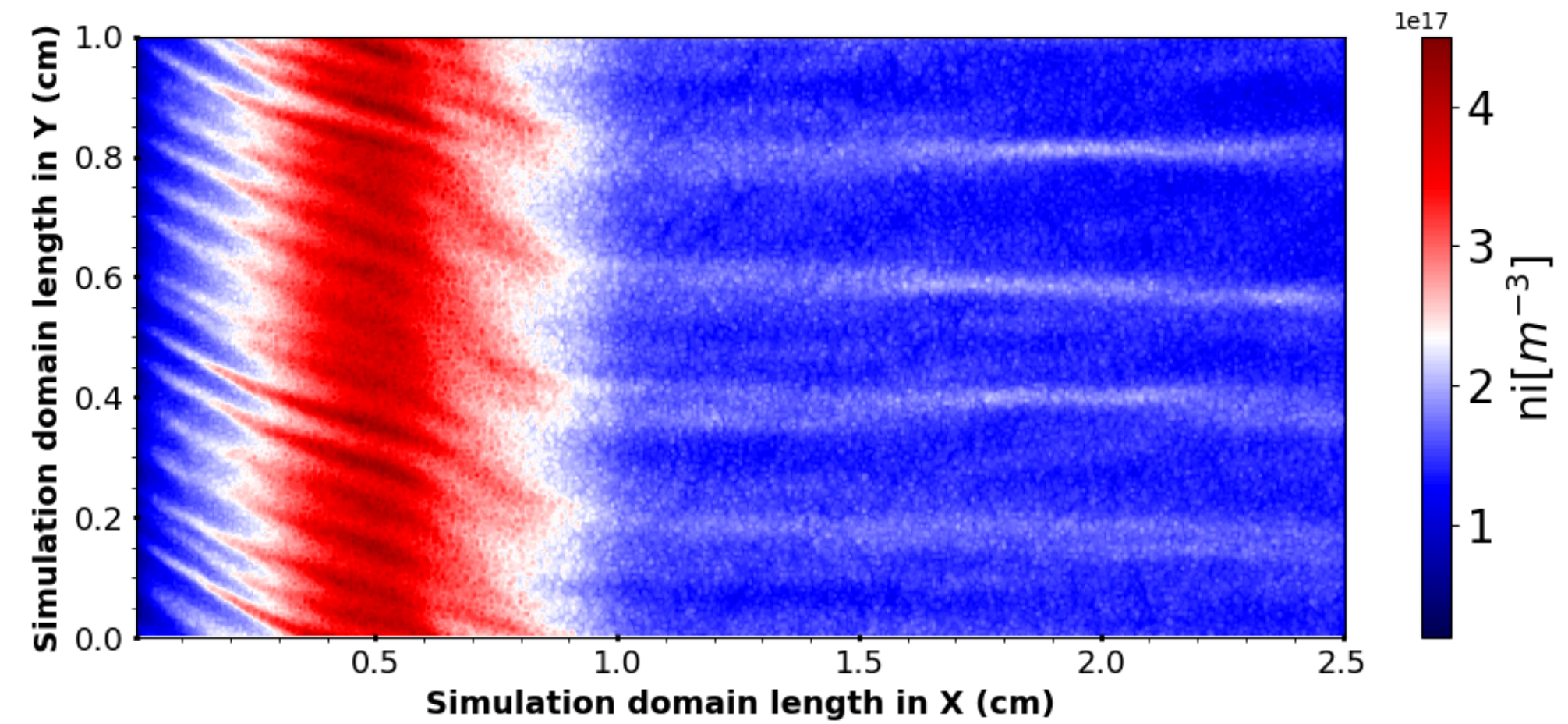}
\caption{ Ion Density at 20 $\mu s$. The image is generated from our in-house 2D-3v  parallel Particle-in-cell code data}  \label{fig:Benchmarking Results}
\end{figure}

\begin{figure}
\includegraphics[width=1.0\textwidth]{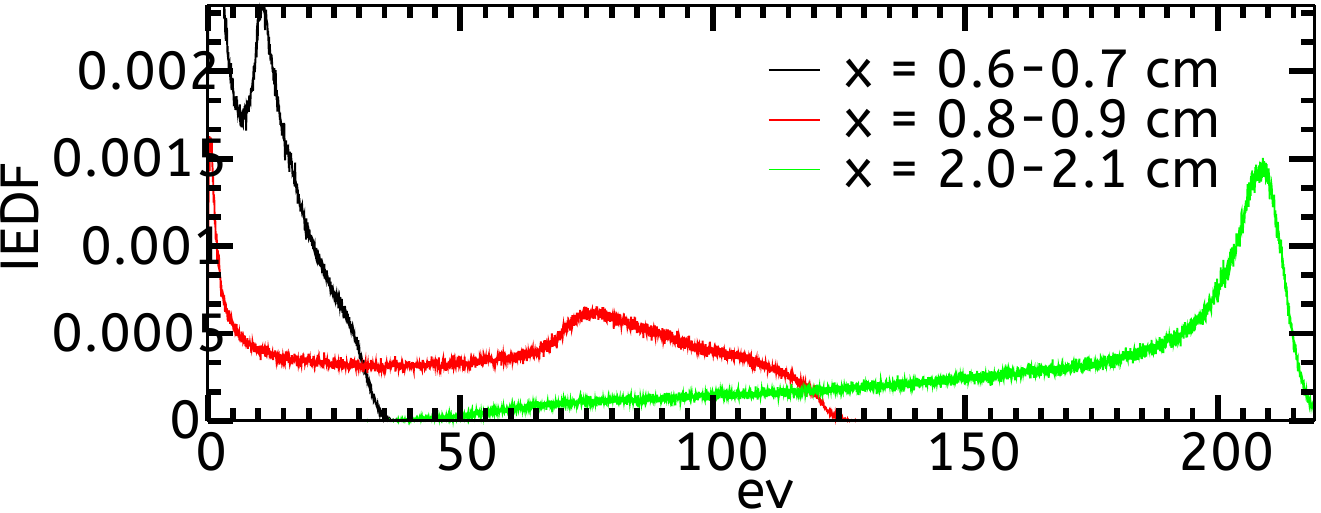}
\caption{ Ion Energy Distribution function at 20 $\mu s$ demarcates the boundary between the ionization and acceleration regions. \cite{boeuf2018b}}  \label{fig:IEDF Benchmarking Results}
\hfill \break
\includegraphics[width=1.0\textwidth]{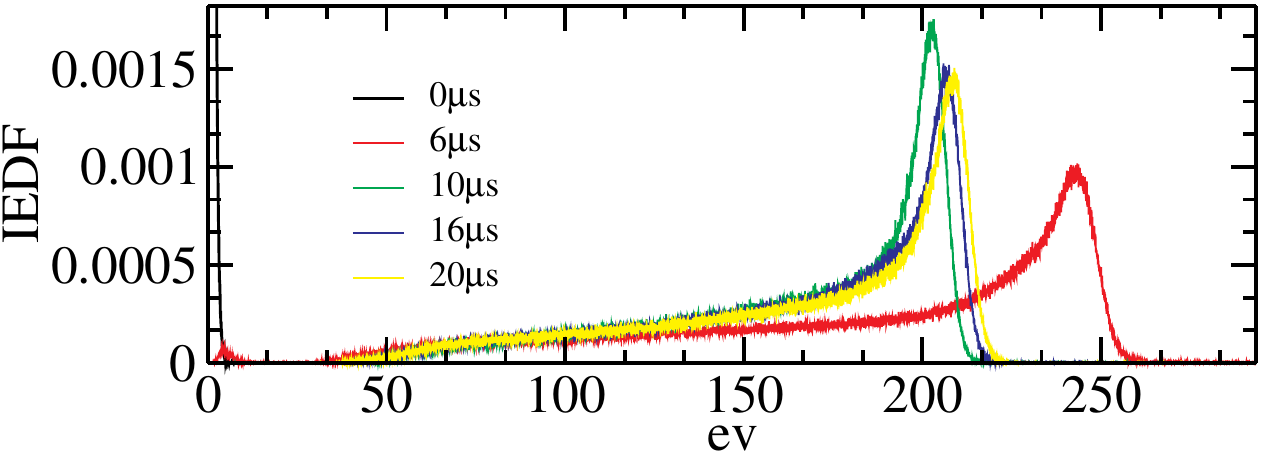}
\caption{ IEDF at x=2.0-2.1cm varying in time \cite{boeuf2018b} indicating that stability is achieved after 10 $\mu s$}
\label{fig:IEDF Benchmarking Results with time}
\end{figure}

Achieving agreement on discharge parameters is required to validate the PIC codes for further analysis of \textit{\textbf{E $\times$ B}} discharges. By utilizing four nodes of the ANTYA HPC and our in-house developed hybrid code \cite{chaudhury2019hybrid}, which fully exploits the computational capabilities of the HPC, the results for the benchmark problem detailed in Section \ref{benchmarking} were obtained in approximately 50,000 core hours. This hybrid approach, which will be explained later, combines the strengths of MPI for inter-node communication and OpenMP for intra-node parallelism, ensuring efficient use of the available computational resources. The instability and the azimuthal wave with large amplitude oscillations of the ion density at $20\mu s$ driven by the large \textit{\textbf{E $\times$ B}} drift is clearly visible in Fig. \ref{fig:Benchmarking Results}. The contour plots illustrate these distributions for a total electron and ion production rate given by the total ion current $J_M=400A/m^2$. The applied voltage is 200 V, driving the wave propagation and accelerating the ions. The ion energy distribution function at 20$\mu s$ for different regions of the simulation domain is shown in Fig. \ref{fig:IEDF Benchmarking Results}. An increase in ion energy from the ionization region to the acceleration region is observed.  Fig. \ref{fig:IEDF Benchmarking Results with time} shows the change in ion energy as the simulation progresses at a fixed region. Results from our in-house developed 2D-3v PIC code have a good qualitative and quantitative match with the results of international benchmark presented in the works \cite{charoy20192d,boeuf2018b}, thereby effectively validating the accuracy of our code.

\section{Parallelization of PIC code via particle-thread binding}\label{Methodology}

To emphasize the novelty of our approach, we provide a brief review of domain decomposition and particle decomposition, the two most widely used traditional parallelization techniques \cite{derouillat2018smilei,qiang2010particle,wang2019modern} for LTP simulations using PIC. In domain decomposition, the entire simulation domain is divided into chunks of subdomain and it also involves keeping track of particles leaving and entering the subdomains \cite{miller2021dynamic,bird2021vpic}. These subdomains are then assigned to the processing elements (thread/core).  In particle decomposition \cite{kaganovich2020physics,shah2023investigation,hara2020cross,kumar2021effects}, particles are divided among threads while ensuring that each thread receives almost the same number of particles since the particles predominantly bear the computational load. This method is favoured for its straightforward implementation and near-perfect load balancing, facilitating the effective implementation of thread-level shared memory parallelization using OpenMP. Distributing particles equally among threads ensures an almost perfect load balance at all times. However, this approach requires all threads to maintain a private copy of charge density grid, which is a major bottleneck towards achieving good efficiency and scalability of a parallel PIC implementation.

\textit{Charge Deposition} module Fig.\ref{fig:flow chart}  \cite{birdsall1991particle}  employs a linear interpolation scheme to distribute the charge contribution of each particle onto the grid points. Each particle contributes its charge to the surrounding four grid points.  The contribution of a particular particle to the grid is determined by the area fraction, as shown in Fig. \ref{fig:conventionalCD} b). These grid points are determined by the position of the particle within the simulation domain. Here $P_1$, $P2$, $P3$, and $P4$ represent the index of the bottom left, bottom right, top left, and top right corners of the cell on the global grid array as shown in Fig. \ref{fig:conventionalCD} b). Many particles will be associated with the same index in the global grid data structure, which is not a problem when \textit{CD} is performed sequentially on particles. However, when done in parallel, multiple threads write to the same location, and race conditions arise. Different methods are used to mitigate this, as elaborated in the section below. 


\subsection{Conventional Parallel CD}
\begin{figure}
    \includegraphics[width=1.0\textwidth]{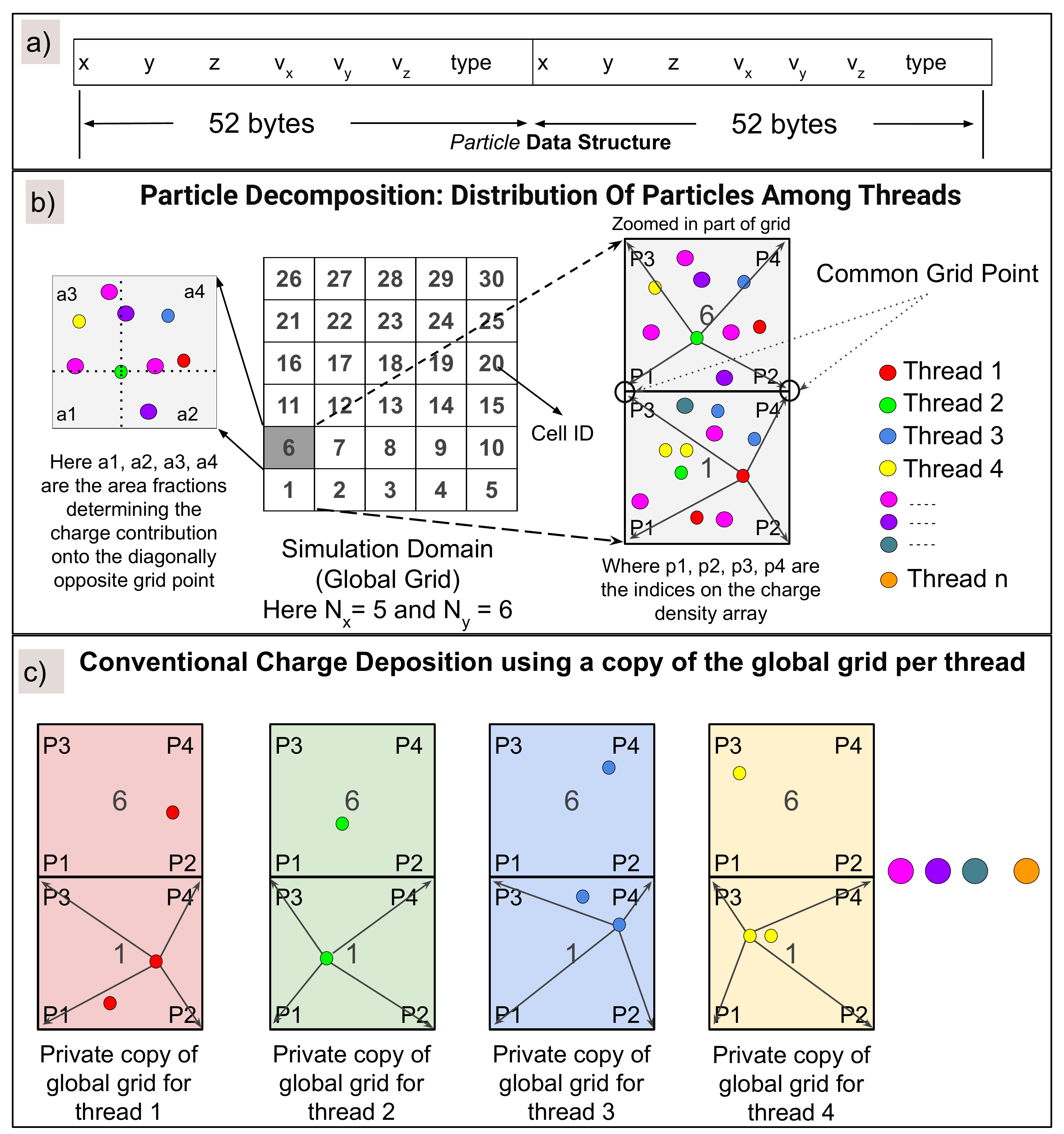}
    \caption{a) The particle data structure, which shows adjacent particles in memory, does not guarantee that the particles are ordered in any specific way. b) Parallelization using particle decomposition on the global grid leads to race conditions at common grid points.  c) Conventional parallel charge deposition uses multiple threads writing charge density into its private grid.}
    \label{fig:conventionalCD}
\end{figure}

In a conventional particle-decomposition strategy (thread-level shared memory parallelization), a simple static scheduling of OpenMP-based parallel \textit{for} can be used to divide the \textit{Particle} array to pick particles iteratively. However, in the context of Particle-to-Grid interpolations utilized in this module, race conditions may arise when two different particles handled by separate threads attempt to update the grid quantities of the same grid point simultaneously. For instance, this occurs at the common vertex illustrated in Fig. \ref{fig:conventionalCD} b) (where $P1$ for cell 6 corresponds to $P3$ for cell 1 and $P2$ for cell 6 corresponds to $P4$ for cell 1). In order to mitigate this problem, conventional techniques involve creating a private copy of the global grid for each thread to address these race conditions\cite{chaudhury2019hybrid}. As Fig. \ref{fig:conventionalCD} c) illustrates, each thread operates independently on its designated particles without requiring any critical region or locks\cite{chaudhury2019hybrid,deluzet2023efficient,miller2021dynamic}. Each private copy is stored in shared memory but is exclusively accessible to its associated thread. After completing charge deposition on their private grids, the master thread combines all private grids to update the final values on the global grid.
While this strategy effectively manages race conditions, it does introduce memory and computing overhead due to the generation and combination of private grids. Further details about this approach can be found in \cite{chaudhury2019hybrid}. The memory overhead associated with maintaining multiple private grids scales according to Eq. \ref{Memory_overhead}:
\begin{equation}
    Memory \propto  NTHR * GRID\_DIMENSIONS
    \label{Memory_overhead}
\end{equation}
Here, $NTHR$ represents the number of threads used, and $GRID\_DIMENSIONS$ is the number of grid points in the Eulerian Grid. The pseudo-code for conventional \textit{CD} is shown below.

\begin{lstlisting}[style=CStyle]
void chargeDeposition()
    double *private_grid_qty = malloc(NTHR * GRID_DIMENSIONS);
    /* Parallelization using openMP for */
    #pragma omp parallel for num_thread(NTHR)
    for(i = 0; i < total_particles; i++)
        int id = omp_get_thread_num();
        int P1,P2,P3,P4; 
        /* calculate the index of the 4 corners of the cell in which this particle resides */
        double update1, update2, update3, update4; /* Updates */
        /* Calculate the value of updates */
        /* Update the private grid quantity's value at nearby grid points */
        private_grid_qty[id * GRID_DIMENSIONS + P1] += update1;
        private_grid_qty[id * GRID_DIMENSIONS + P2] += update2;
        private_grid_qty[id * GRID_DIMENSIONS + P3] += update3;
        private_grid_qty[id * GRID_DIMENSIONS + P4] += update4;
    /* Combine the private grids to form the updated global (Synchronization) */
    for(i = 0; i < NTHR; i++)
        for(j = 0; j < GRID_DIMENSIONS; j++)
            global_grid[j] = private_grid_qty[i * GRID_DIMENSIONS + j];
\end{lstlisting}

\subsection{Proposed particle-thread binding parallel interpolation of charge density}

\begin{figure}
\centering    \includegraphics[width=0.85\textwidth]{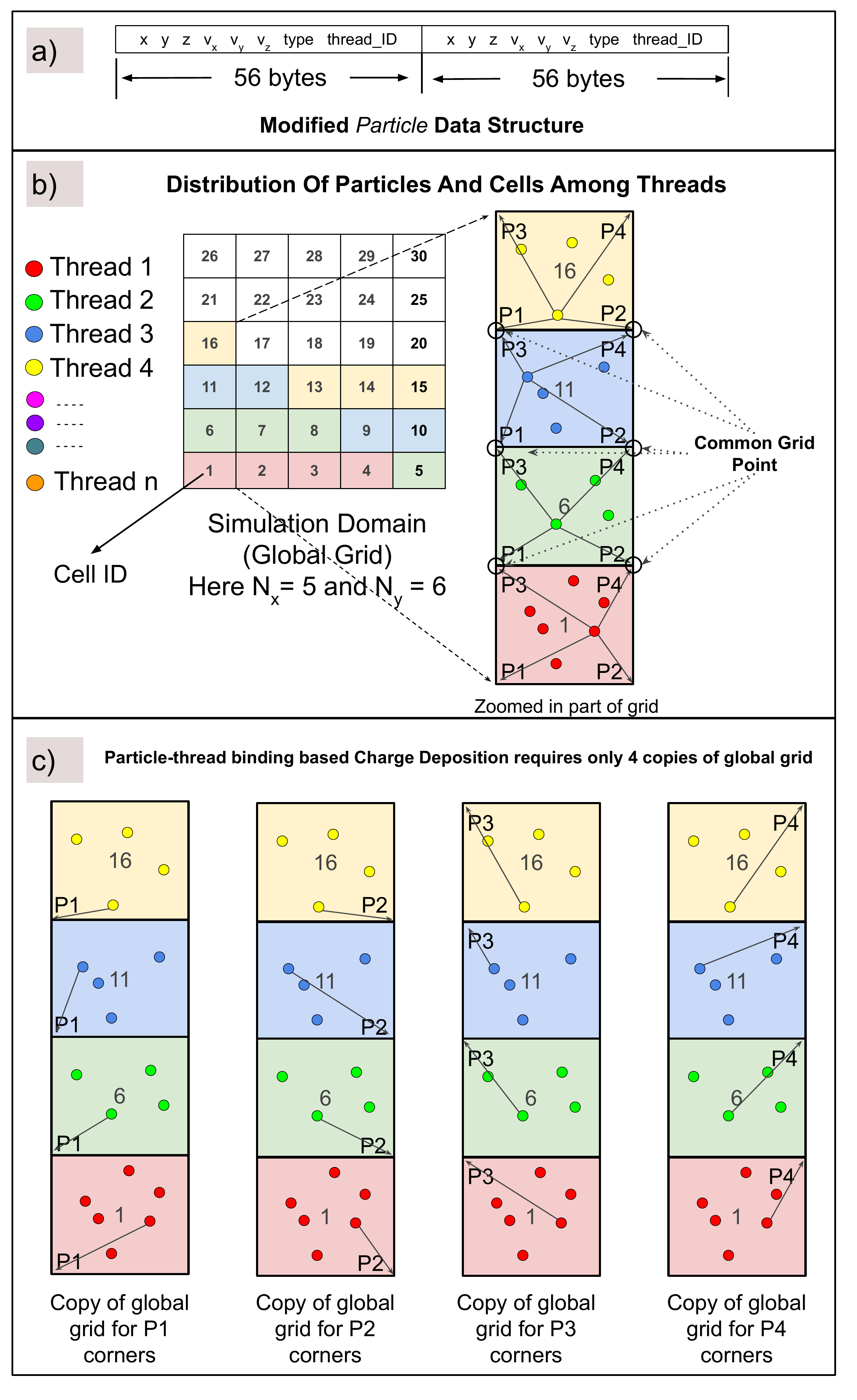}
    \caption{a) Modified Data Structure enables the allocation of \textit{thread\_ID}, effectively binding particles to thread. b) The Mapping Function calculates the \textit{thread\_ID} using the \textit{cell\_ID}. This ensures that all the particles in a cell will be bound to the same thread. c)  To mitigate race conditions during charge deposition at common grid points, four copies of the global grid(one for each corner of a cell) are used. }
    \label{fig:P1P2P3P4}
\end{figure}

Conventional \textit{CD} module encounters scalability challenges as its overhead scales with \textit{NTHR}, leading to diminishing speedup from the additional threads. To mitigate this, we propose a PTB charge deposition with fixed overhead. Implementing this requires \textit{Particle} data structure to be modified to accommodate one additional integer to store \textit{thread\_ID} see Fig. \ref{fig:P1P2P3P4} a). During the initialization of the simulation setup, a two-step mapping function calculates the \textit{cell\_ID} \cite{Hur_2019} based on the particle's location in the simulation domain. Using the \textit{cell\_ID}, the mapping function then calculates a \textit{thread\_ID}, which is stored in the \textit{Particle} data structure. \textit{thread\_ID} effectively binds particles to a thread. Depending on the distribution, the mapping function can be dynamic or static to balance load. Since the density is generally uniform in the $y$ direction in the simulation setup of the benchmarking problem, load balancing can be achieved straightforwardly by dividing the rows of cells almost equally among the allocated threads. The scheme is illustrated in Fig. \ref{fig:P1P2P3P4} b). This ensures that particles in a particular cell are always mapped to the same thread.

This approach requires only four private grids for charge-density computation, regardless of the number of threads used. These four private grids are exact copies of the global grid, which stores a partial sum for specific corners identified as P1, P2, P3, and P4; e.g., every cell has a P1 corner, the P1 private grid will store only the contribution of the charge density on the P1 corner of all the cells. Similarly, the contributions associated with remaining three corners (P2,P3,P4) are stored in three different private grids (labelled as P2 - P4). This approach uses the shared memory concept; thus, these private grids are entirely accessible by all concurrent threads. Race conditions arise only when different threads pick particles from the same cell, which cannot be guaranteed in particle decomposition. To mitigate this, most previous implementations  \cite{MERTMANN20112161} \cite{Stantchev2008} required sorting at every step. Sorting the particle array will ensure that no particle allocated to another thread will lie between the start and end index of the particle array assigned to a thread. Sorting the entire \textit{Particle} array incurs significant computational overhead. To avoid sorting, the proposed approach uses the \textit{thread\_ID} stored in the \textit{Particle} data structure. Using OpenMP, multiple threads are spawned. All these threads traverse the entire array of \textit{Particle} while initiating a flag-like mechanism of a boolean check at the beginning to match the \textit{thread\_ID} stored in \textit{Particle} with the current OpenMP thread id. If a match is found, proceed with interpolation; otherwise, pick the next particle. In the subsequent iterations, the \textit{Mover} module contains the mapping function, which will update the \textit{thread\_ID} based on the new location of the particle, as explained in section \ref{Flag-Based Mover}. This ensures that no two threads get particles from the same cell, and each thread performs charge distribution of particles assigned to it. The boolean check requires minimal computation compared to the full sorting of the particle arrays. After completing charge deposition on the private grids, combine all private grids to update the final values on the global grid since these grids carry partial sums of each corner, as shown in Fig. \ref{fig:P1P2P3P4} c). Here, the two key advantages are the minimal changes required to the conventional method to implement this method and requirement of only 4 private grids irrespective of number of cores. The pseudocode for this implementation is given below.

\begin{lstlisting}[style=CStyle]
void chargeDeposition()
    double *private_grid_qty = malloc(4 * GRID_DIMENSIONS);
    /* Spawns multiple threads */
    #pragma omp parallel num_thread(NTHR)
    for(i = 0; i < total_particles; i++)
        int id = omp_get_thread_num();
        /* Skip particles not assigned to this thread */
        if(particles_array[i].thread_id!=id) 
            continue 
        int P1,P2,P3,P4; 
        /* calculate the index of the 4 corners of the cell in which this particle resides */
        double update1, update2, update3, update4; /* Updates */
        /* Calculate the value of updates */
        /* Update the private grid quantity's value at nearby grid points */
        private_grid_qty[0 * GRID_DIMENSIONS + P1] += update1;
        private_grid_qty[1 * GRID_DIMENSIONS + P2] += update2;
        private_grid_qty[2 * GRID_DIMENSIONS + P3] += update3;
        private_grid_qty[3 * GRID_DIMENSIONS + P4] += update4;
    /* Combine the private grids to form the updated global (Synchronization) */
    for(i = 0; i < 4; i++)
        for(j = 0; j < GRID_DIMENSIONS; j++)
            global_grid[j] = private_grid_qty[i * GRID_DIMENSIONS + j];
\end{lstlisting}

\subsection{Mover}
\label{Flag-Based Mover}
The \textit{Mover} module is easy to parallelize by distributing particles evenly among the threads. Here, interpolations occur from the grid to the particle. As particles are allocated among threads, each thread updates a particular particle's phase space information just once by reading grid quantities. Consequently, parallel updates avoid race conditions as each thread exclusively does writing, and multiple threads can simultaneously read grid quantities, making this process embarrassingly parallel. Additionally, equal distribution of particles ensures efficient load balancing, facilitating simple implementation with static scheduling. After the first iteration, the mapping function computes the \textit{cell\_ID} and \textit{thread\_ID} in the \textit{Mover} itself, to be used in the next iteration of \textit{CD}. It minimally impacts the arithmetic intensity of the \textit{Mover} since the new location of the particle is already available in the cache. The pseudo-code for the \textit{Mover} module is shown below.
\begin{lstlisting}[style=CStyle]
void mover()
    /* Parallelization over particles using private grids */
    #pragma omp parallel for
    for(i = 0; i < total_particles; i++)
        double updateX, updateY; /* Updates in position */
        double updateVx, updateVy, updateVz; /* Updates in velocities */
        /* Calculate the value of updates based on the value of 
        grid quantities at neighboring grid points */
        /* Update the phase space information at particle_array[i] */
        particles_array[i].x += updateX; 
        particles_array[i].y += updateY;
        particles_array[i].vx += updateVx; 
        particles_array[i].vy += updateVy;
        particles_array[i].vz += updateVz;

        int cell_ID,thread_ID;
        /* Calculate Cell_ID and thread_ID */
	particles_array[i].thread_id=thread_ID;
\end{lstlisting}


\subsection{Hybrid Parallelization Strategy}
For HPC platforms, hybrid parallelization utilizes multiple interconnected nodes that communicate using the MPI library (Node-level parallelism). Each node possesses a specific number of cores and memory capable of launching hardware thread (thread-level parallelism). In the hybrid parallelization on an HPC system, particles are initially divided among nodes, with each node executing the \textit{CD} and \textit{Mover} modules for its assigned particles independently. Since particle-particle interaction is mediated through the grids, a common view of the grid needs to be kept, which is facilitated through MPI communications. Nodes maintain private copies of grids with partial sums of quantites. Each node utilizes thread-level parallelization to update node-level grid quantities. Global aggregation involves communication between nodes, using the \textit{MPI\_Allreduce} function on the grid quantities to optimize costly network communication and MPI Barrier to synchronize processes. The communication overhead associated with particle decomposition stems from the need to communicate the grid quantities over a network and sum them up. Load balancing is automatically handled through particle decomposition, which is done at the beginning of the simulation. Table \ref{tab: Memory complexity} provides the details of the memory complexity associated with different implementations of \textit{CD}.

\begin{table}[]
    \centering
    \begin{tabular}{|c|c|c|c|c|c|c|}
        \hline
        \textbf{Config} &
        \multicolumn{2}{|c|}{\textbf{250x100}} &
        \multicolumn{2}{|c|}{\textbf{500x200}} &
        \multicolumn{2}{|c|}{\textbf{1000x400}} \\
        \hline
        \textbf{} &
        \textbf{Conv} & \textbf{PTB-PIC} &
        \textbf{Conv} &
        \textbf{PTB-PIC} &
        \textbf{Conv} &
        \textbf{PTB-PIC} \\
        \hline
        1N 1C & 390 KB & 1.5 MB & 1.5 MB & 6 MB & 6 MB & 24 MB\\
        \hline
        1N 2C & 781 KB &  1.5 MB & 3 MB & 6 MB & 12 MB & 24 MB\\
        \hline
        1N 4C & 1.5 MB & 1.5 MB &6 MB &6 MB  &24 MB & 24 MB\\
        \hline
        1N 8C & 3 MB & 1.5 MB&12 MB & 6 MB &48 MB & 24 MB\\
        \hline
        1N 40C & 15 MB & 1.5 MB & 60 MB& 6 MB & 240 MB & 24 MB\\
        \hline
        2N 40C & 30 MB & 3.0 MB &120 MB & 12 MB &480 MB & 48 MB\\
        \hline
        4N 40C & 60 MB & 6 MB &240 MB& 24 MB &960 MB & 96 MB\\
        \hline
        10N 40C & 150 MB & 15 MB &600 MB & 60 MB & 2.4 GB& 240 MB\\
        \hline
        25N 40C & 375 MB & 37.5 MB&1.5 GB & 150 MB &6 GB & 600 MB\\
        \hline
    \end{tabular}
    \caption{{Memory complexity in \textit{CD} based on Hybrid parallelization for both algorithm. Here, N indicates the number of nodes and C indicates cores. The configurations taken here are based on the ANTYA HPC at IPR Gandhinagar. The grid sizes (first row) are the typical scenarios encountered in \textit{\textbf{E $\times$ B}} plasma simulations \cite{boeuf2018b}. }}
    \label{tab: Memory complexity}
\end{table}

\section{Results and Discussion}\label{Results}

In this section, we present a comprehensive analysis of the scalability and performance of Electrostatic Particle-In-Cell simulations, utilizing both the conventional \textit{CD} module and a PTB \textit{CD} module on shared memory systems and on an HPC cluster. Our experiments encompass typical scenarios encountered in \textit{\textbf{E $\times$ B}} plasma simulations \cite{boeuf2018b}. Five different configurations have been selected, considering the varying computational loads of the particle and grid data structures (see Table. \ref{tab: configuration}). Speedups and runtimes are sensitive to the total number of particles, the number of particles per cell (PPC) and grid dimensions, which play a crucial role in determining the final runtime. Thus, a combination of small and large grids with high and low ppc counts is used. The simulation parameters of these configurations closely resemble the benchmark problem mentioned in the previous section. All simulations are run for 2000 iterations multiple times, and average runtimes are noted for comparison. A serial code implies that the code was run on a single core of the corresponding machine. Details of the physical parameters can be found in the following work \cite{boeuf2018b}.

\begin{table}[htbp]
    \centering
    \caption{Typical configurations encountered in \textbf{\textit{E $\times$ B}} plasma simulations}
    \begin{tabular}{|c|c|c|}
        \hline
        \textbf{Grid Dimensions} & \textbf{ppc} & \textbf{Total particle in millions} \\
        \hline
        250 x 100 & 18 & 0.9 \\
        \hline
        250 x 100 & 100 & 5  \\
        \hline
        500 x 200 & 18 & 3.6  \\
        \hline
        500 x 200 & 100 & 20  \\
        \hline
        1000 x 400 & 18 & 14  \\
        \hline
    \end{tabular}
    \label{tab: configuration}
\end{table}


\begin{figure}
    \centering
    \includegraphics[width=\textwidth]{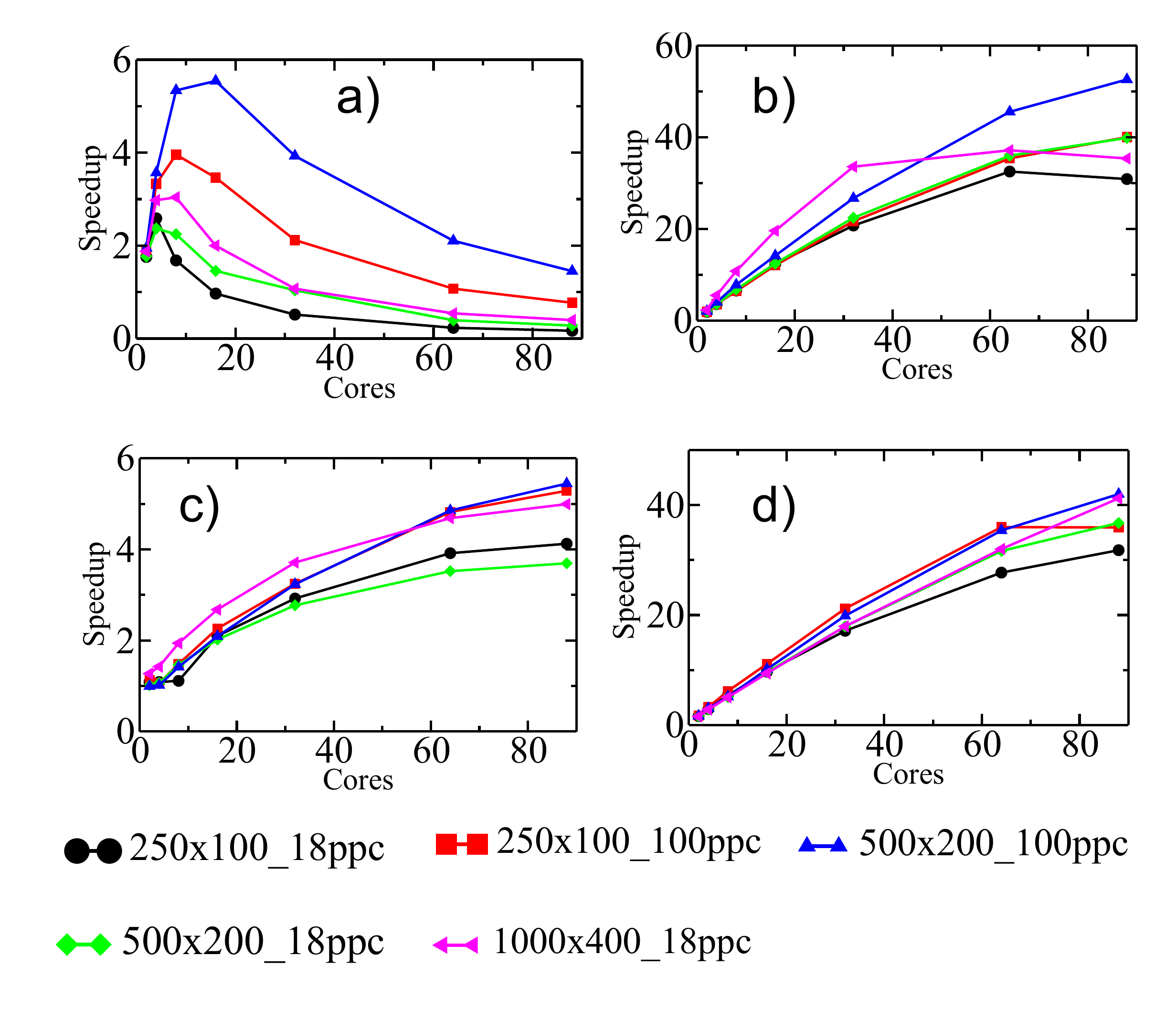}
    \caption{The speedup of individual modules on a shared memory system. a) Speedup of conventional \textit{CD} compared to its serial implementation. b) Speedup of conventional \textit{Mover} compared to its serial implementation. c) The speedup of PTB \textit{CD} compared to its serial implementation. d) Speedup of PTB \textit{Mover} compared to its serial implementation}
    \label{fig:shared CD mover combined}
\end{figure}
\subsection{Speedup studies of PIC on shared memory architecture}
The architecture used in this work features a 4-socket system, with each socket housing an Intel(R) Xeon(R) Gold 6238 CPU operating at a base clock frequency of 2.10 GHz. Each Intel CPU comprises 22 cores, resulting in a cumulative system core count of 88. These CPUs are interconnected via a maximum of 3 UPI links, enabling seamless communication among the four CPUs and effectively allowing them to function as a single processor with 88 cores and 121 MB of last-level cache shared among all cores. Our specific system setup has 512 GB of RAM. 

We investigate the impact of different configurations (Table. \ref{tab: configuration}) on the speedup achieved and provide insights into the scalability of the proposed PTB method. We begin by analyzing the scalability of conventional PIC simulations employing the conventional \textit{CD} module on a shared memory architecture. 
A critical observation from our study is that increasing the number of cores does not result in a linear increment in speedup. To investigate the problem, we looked at the scalability of individual modules. Fig. \ref{fig:shared CD mover combined} a) illustrates the drawback associated with conventional \textit{CD}. We observe a diminishing speedup beyond a certain number of cores for \textit{CD} ( Fig. \ref{fig:shared CD mover combined}). This is attributed to the overhead of creating private grids equal to the number of threads. Additional memory is required to keep all the grids in the shared cache for cache hits. However, as number of private grids increases, due to limited size of cache, only a portion of the grid is stored in the cache, resulting in significant cache misses. There is also an increase in the computational overhead associated with creating and summing all the private grids. Diverse configurations display varied peaks of possible speedup at distinct core counts, reflecting the balance between increased overhead and increased parallelism. Configurations with higher ppc exhibit notable speedup, indicating the effectiveness of harnessing additional cores to manage the increased workload. However, the complex balance between grid size and parallelism determines the speedup achieved. A key observation is that the optimal number of cores for maximum speedup varies with problem size. Fig. \ref{fig:shared CD mover combined} b) illustrates the scalability of the conventional \textit{Mover} module; the trend clearly indicates an almost linear relation between core count and speedup. Insights from these analyses reveal why adding more cores does not always lead to better end-to-end performance of a PIC simulation based on conventional parallelization techniques. The overall speedup relies on how well these two modules scale individually.



Speedup of PTB based \textit{CD} is shown in Fig. \ref{fig:shared CD mover combined} c) showcasing the scalability of the proposed PTB-based algorithm. This can be attributed to the fixed overhead associated with the method (due to only 4 private grids), thereby allowing parallelism to increase as more threads are employed, enabling the utilization of all available cores. A key observation here is that maximum performance is attained when all cores are deployed. Consequently, end users no longer need to conduct performance testing before execution to determine the optimal number of cores, as in plasma simulations, the ppc changes dynamically during runtime, but PTB-based ES-PIC consistently operates at the optimal point. Speedup achieved depends on configuration.


    


\subsection{Scalability studies of particle-thread binding based PIC on a distributed system}

\begin{figure}
    \centering
    \includegraphics[width=0.95\textwidth]{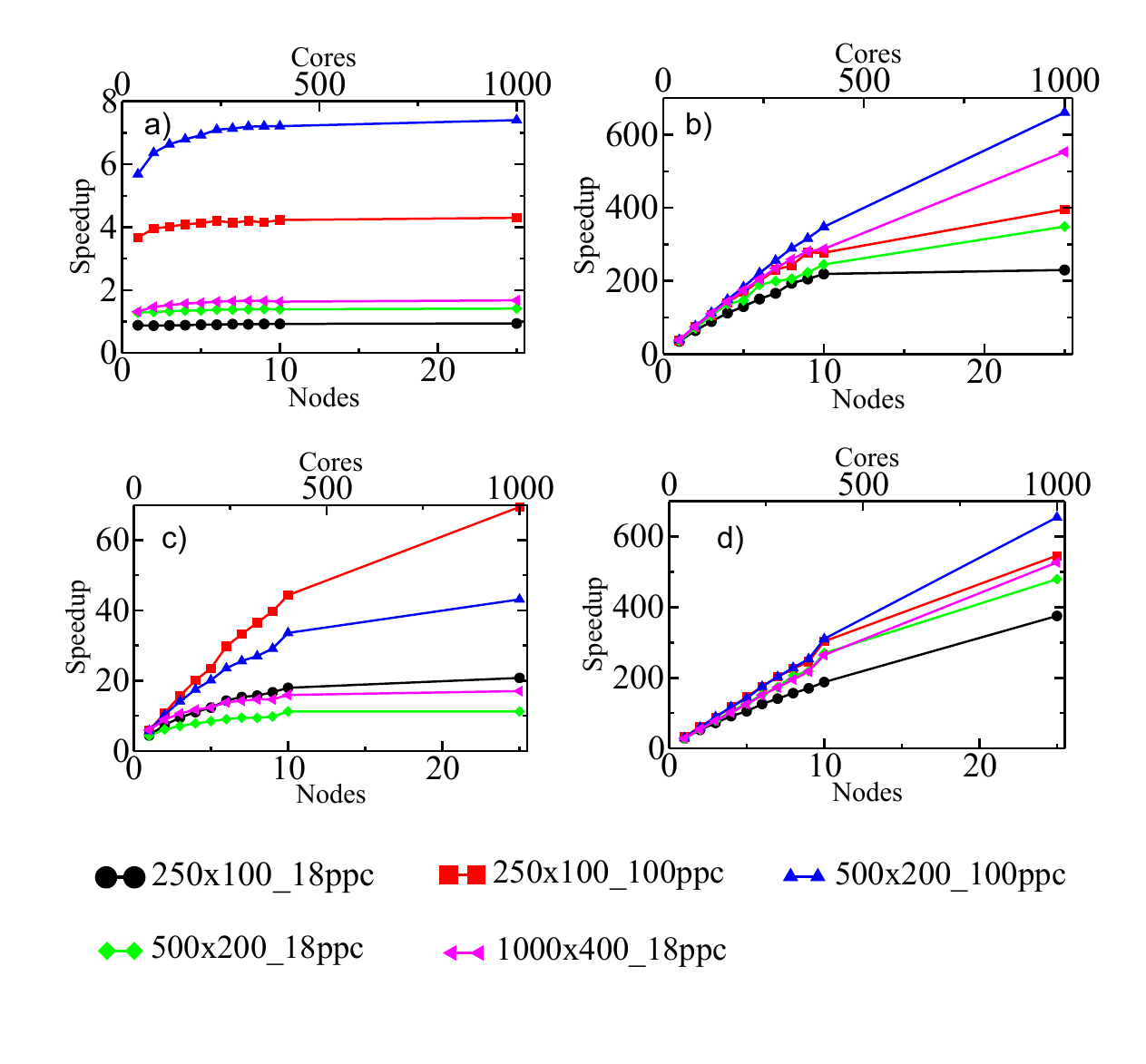}
    \caption{This figure compares the speedup of individual modules on ANTYA HPC. a) Speedup of conventional \textit{CD} compared to its serial implementation. b) Speedup of conventional \textit{Mover} compared to its serial implementation. c) The speedup of PTB \textit{CD} compared to its serial implementation. d) Speedup of PTB based \textit{Mover} compared to its serial implementation.}
    \label{fig:distributed CD mover combined}
    
\end{figure}

To test the scalability \cite{KHAZIEV201887} of the PTB based PIC on a distributed system, we executed the code on the ANTYA HPC facility described in section 2.3 and utilized a maximum of 1000-core to assess the speedup of individual modules. The conventional \textit{CD} exhibits a trend, as shown in Fig. \ref{fig:distributed CD mover combined} a), where using a large number of cores becomes inefficient due to limited scalability, resulting in substantial wastage of computational resources. The speedup is evaluated relative to the serial implementation. In contrast, the PTB \textit{CD} demonstrates a clear advantage, efficiently scaling with an increasing core count, as illustrated in Fig. \ref{fig:distributed CD mover combined} c).
 As expected, the \textit{Mover} Fig. \ref{fig:distributed CD mover combined} b) and d) modules are scalable and show excellent speedup in both codes. The PTB-based \textit{CD} achieves a speedup of over 60× when utilizing 1,000 cores across 25 nodes, significantly outperforming the conventional parallel algorithm, which achieves only a 7× speedup. Notably, this performance trend remains consistent across various problem sizes, demonstrating the robustness of the PTB approach in different problem sizes.

\subsection{End-to-End speedup and its explanation}

\begin{figure}
    \centering
    \includegraphics[width=0.95\textwidth]{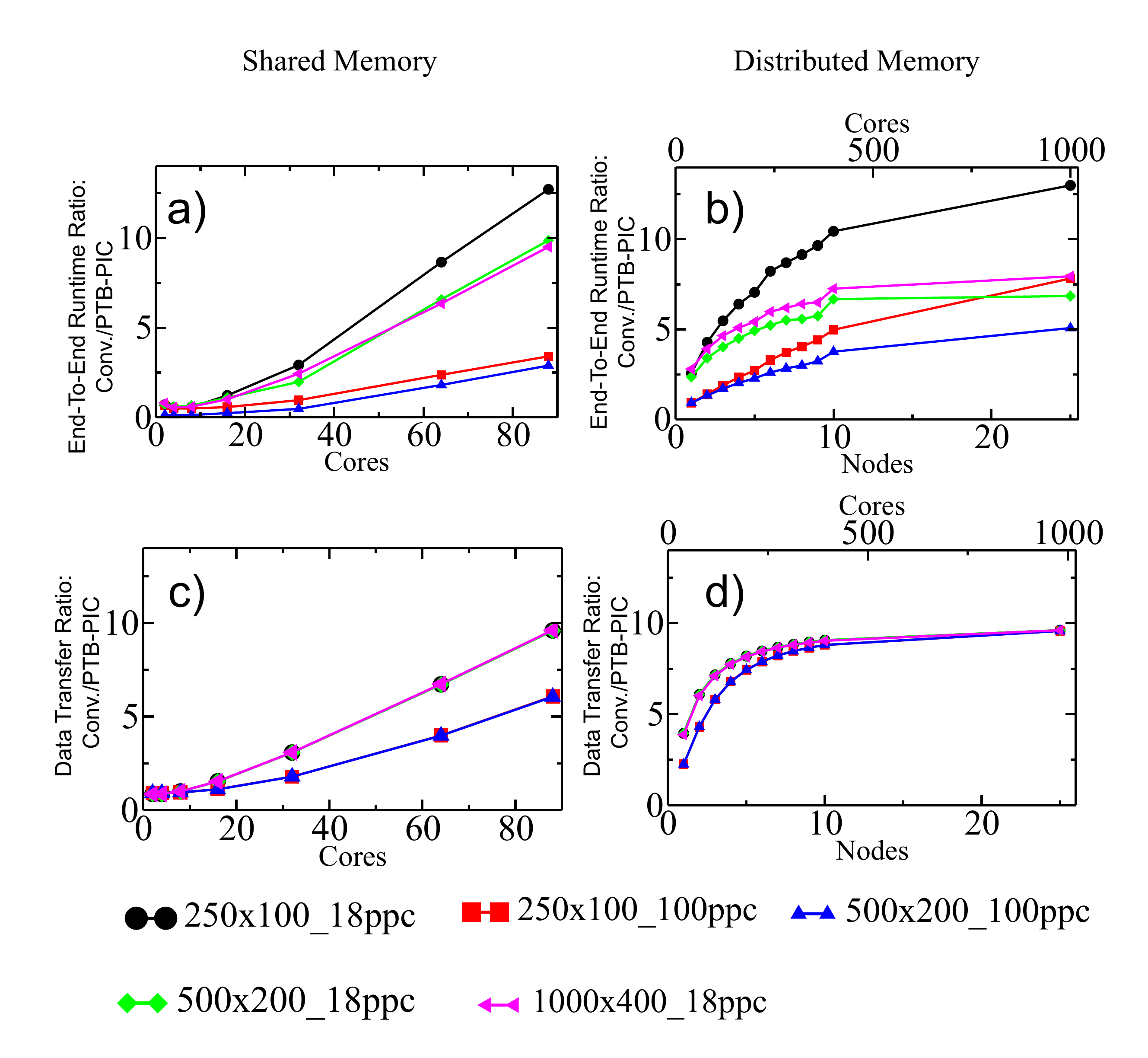}
    \caption{a) The ratio of total runtimes of conventional PIC vs PTB-PIC on a shared memory system. b) Ratio of total runtimes of conventional PIC vs PTB-PIC on a distributed memory system. c) Ratio of total data transfer required for conventional PIC vs PTB-PIC on a shared memory system d) Ratio of total data transfer of conventional PIC vs PTB-PIC on a distributed memory system.}
    \label{fig:End-To-End}
    
\end{figure}

The results presented in the previous section clearly demonstrate that the PTB-based implementation of the PIC algorithm effectively overcomes the scalability limitations of conventional parallel PIC codes. The observed trend suggests the potential for further speedup with additional cores, a capability not achievable with conventional PIC. Fig. \ref{fig:End-To-End}(a) and (b) illustrates this trend in a direct performance comparison between the two approaches on both shared-memory and distributed-memory systems. As computing architectures transition from multi-core to many-core systems, the PTB-based PIC is expected to offer significant benefits over conventional PIC implementations.

To investigate the improved performance, we analyzed the data access pattern of the \textit{CD} module as a function of cores (threads) (see Eq. \ref{eq:codeBalance conventional PIC}). The data access for \textit{CD} is proportional to the combination of \textit{Particle} and \textit{Grid} components, where $N$ is the number of particles, $Core$ is the number of cores used in shared memory system and cores per $Node$ for distributed memory system, $N_G$ is the number of grid points and $Nodes$ is the number of nodes used. In both Eq. \ref{eq:codeBalance conventional PIC} and Eq. \ref{eq:codeBalance flag based PIC}, 
the primary computational task is parallel particle-to-grid data interpolation, where $\frac{N}{Core * Nodes} + N_G$ is the data required per core for interpolation using the conventional method as particle decomposition distributes the particle data structure equally, and a private grid is assigned to each processing core; however, the performance will strongly depend on the efficiency of data access through the different data paths involving the memory hierarchy. More cache hits will naturally lead to better performance, making it imperative to keep all the \textit{Grid} data in the cache. Once interpolation is done, the reduction operation requires $Core*N_G*Nodes$ amount of data to sum the private grid, which is scaled by the number of cores. In PTB-based \textit{CD} $\frac{N}{Core * Nodes} + 4N_G$ is the data required for interpolation, and $4*N_G*Nodes$ is the data required for reduction.
\begin{equation}
Data(Conventional) \propto \frac{N}{Core * Nodes} +
N_G
+ Core*N_G*Nodes
    \label{eq:codeBalance conventional PIC}
\end{equation}
\begin{equation}
Data(PTB) \propto \frac{N}{Core * Nodes} + 4N_G + 4N_G*Nodes
    \label{eq:codeBalance flag based PIC}
\end{equation}
The ratio of Eq. \ref{eq:codeBalance conventional PIC} to Eq. \ref{eq:codeBalance flag based PIC} is denoted as the data transfer ratio. Fig. \ref{fig:End-To-End} c) shows the data transfer ratio with \textit{Nodes} set to 1 as it is a shared memory system, it aligns well with the end-to-end runtime ratio of conventional PIC and PTB-PIC shown in Fig. \ref{fig:End-To-End} a). The $250\times100\_18ppc$ case is a notable exception due to its small size, allowing both \textit{Particle} and \textit{Grid} data to fit entirely in the L3 cache in the case of PTB, resulting in exceptional speedup. This trend is also visible in the distributed memory system (see Fig. \ref{fig:End-To-End} b) and d) ). Furthermore, the strong agreement between the speedup ratio and the data transfer ratio in Fig. \ref{fig:End-To-End} suggests that the proposed method exhibits minimal dependence on the underlying hardware. A direct runtime comparison reveals a performance gain in the range of 2.8-12X for the shared memory systems and 5-13X for the distributed memory systems.

These results demonstrate that the PTB-based approach significantly reduces computational time by minimizing data transfer overhead, a critical factor for efficiency in large-scale distributed computing, essentially making it hardware-independent. Consequently, it can provide a performance boost to simulations of various problem sizes in both shared and distributed memory environments. The load balance achieved by particle decomposition further enhances scalability by reducing the computational load on each \textit{Node} as their number increases, enabling greater speedups compared to conventional PIC codes. The problem size \textit{$500\times200\_100ppc$} simulated here closely resembles the 2D axial-azimuthal particle-in-cell benchmark for low-temperature
partially magnetized plasmas \cite{charoy20192d}, which requires 50000 core hours using the conventional approach on our HPC system, whereas with the PTB-based \textit{CD}, we can observe a reduction in simulation time by a factor of 5 or a speedup of 5X. Overall, PTB-based approach offers excellent scalability and resource efficiency across various problem sizes, making it a highly effective solution for kinetic simulation of LTPs using PIC.

\section{Conclusion}
Scalability is a challenge in massively parallel Particle-in-Cell (PIC) codes when executed on large-scale distributed HPC platforms and shared-memory workstations. This limitation primarily stems from the conventional parallel private-grid-based charge deposition (CD) scheme, which scales poorly with core count as it requires a separate private grid for each thread, resulting in significant memory overhead and increased computational costs. To address this, we introduce a novel Particle Thread Binding (PTB) scheme, which significantly reduces the memory overhead by requiring only four private copies for CD in shared-memory systems and four per node in distributed-memory systems. To provide a comprehensive understanding, we first describe our in-house developed Hybrid (OpenMP + MPI) 2D-3v massively parallel PIC code, which follows the conventional private grid-based PIC approach and details its implementation and associated performance bottlenecks. Subsequently, we describe the PTB-based PIC implementation. We also validate the numerical accuracy of the PIC implementations by benchmarking it against a recently published study focused on benchmarking PIC codes for low-temperature plasmas in a \textit{\textbf{E $\times$ B}} discharge\cite{charoy20192d}. We demonstrate a strong quantitative agreement between our results and those of the benchmark simulations by the comparative analysis of the electron energy distribution and the ion density plots with those published in prior studies. Several case studies with different grid sizes and particles per cell demonstrate that the proposed PTB-based PIC significantly improves performance over conventional PIC by effectively mitigating the scalability issue associated with the conventional CD. We observed an end-to-end speedup of 2.8-12X for shared memory systems and 5-13X for distributed memory HPC systems when using PTB-PIC, compared to conventional PIC. We anticipate that this efficient PTB-based parallelization of PIC codes will significantly reduce runtimes for these resource-intensive low-temperature plasma simulations.

\textbf{Acknowledgements}\\
This work has been carried out using the HPC facilities at DA-IICT, IPR Gandhinagar and IIT Kanpur. We acknowledge National Supercomputing Mission (NSM), Government of India, for providing computing
resources of ‘PARAM Sanganak’ at IIT Kanpur. Special thanks to Jayram Ashok for his assistance.
This work was supported by the Department of Science
and Technology, Government
of India.\\

\textbf{Credit authorship contribution statement}\\

Libin Varghese: Writing – original draft (lead), Conceptualization, Validation, Investigation, Software (lead), Methodology, Data curation, Data visualization. \\
Bhaskar Chaudhury: Writing – original draft, Methodology, Conceptualization, Formal Analysis, Supervision, Project administration, Resources, Funding acquisition (lead). \\
Miral Shah: Software, Methodology. \\
Mainak Bandyopadhyay: Resources, Supervision.\\

\textbf{Declaration of Competing Interest}
The authors declare that they have no known competing financial interests or personal relationships that could have appeared to influence the work reported in this paper.

\vspace{20pt}
\bibliographystyle{abbrv}
\bibliography{article}

\providecommand{\newblock}{}
\begin{thebibliography}{10}
\expandafter\ifx\csname url\endcsname\relax
  \def\url#1{{\tt #1}}\fi
\expandafter\ifx\csname urlprefix\endcsname\relax\def\urlprefix{URL }\fi
\providecommand{\eprint}[2][]{\url{#2}}

\bibitem{adamovich20222022}
Adamovich I, Agarwal S, Ahedo E, Alves L~L, Baalrud S, Babaeva N, Bogaerts A, Bourdon A, Bruggeman P, Canal C {\em et~al.\/} 2022 {\em Journal of Physics D: Applied Physics\/} {\bf 55} 373001

\bibitem{laroussi2017perspective}
Laroussi M, Lu X and Keidar M 2017 {\em Journal of Applied Physics\/} {\bf 122}

\bibitem{ovanesyan2019atomic}
Ovanesyan R~A, Filatova E~A, Elliott S~D, Hausmann D~M, Smith D~C and Agarwal S 2019 {\em Journal of Vacuum Science \& Technology A\/} {\bf 37}

\bibitem{levchenko2020perspectives}
Levchenko I, Xu S, Mazouffre S, Lev D, Pedrini D, Goebel D, Garrigues L, Taccogna F and Bazaka K 2020 {\em Physics of Plasmas\/} {\bf 27}

\bibitem{speth2006overview}
Speth E, Falter H, Franzen P, Fantz U, Bandyopadhyay M, Christ S, Encheva A, Fr{\"o}schle M, Holtum D, Heinemann B {\em et~al.\/} 2006 {\em Nuclear Fusion\/} {\bf 46} S220

\bibitem{turner2017computer}
Turner M~M 2017 {\em Plasma Processes and Polymers\/} {\bf 14} 1600121

\bibitem{taccogna2019latest}
Taccogna F and Garrigues L 2019 {\em Reviews of Modern Plasma Physics\/} {\bf 3} 12

\bibitem{kim2005particle}
Kim H, Iza F, Yang S, Radmilovi{\'c}-Radjenovi{\'c} M and Lee J 2005 {\em Journal of Physics D: Applied Physics\/} {\bf 38} R283

\bibitem{von2017foundations}
Von~Keudell A and Schulz-Von Der~Gathen V 2017 {\em Plasma Sources Science and Technology\/} {\bf 26} 113001

\bibitem{Adamovich_2017}
Adamovich I, Baalrud S~D, Bogaerts A, Bruggeman P~J, Cappelli M, Colombo V, Czarnetzki U, Ebert U, Eden J~G, Favia P, Graves D~B, Hamaguchi S, Hieftje G, Hori M, Kaganovich I~D, Kortshagen U, Kushner M~J, Mason N~J, Mazouffre S, Thagard S~M, Metelmann H~R, Mizuno A, Moreau E, Murphy A~B, Niemira B~A, Oehrlein G~S, Petrovic Z~L, Pitchford L~C, Pu Y~K, Rauf S, Sakai O, Samukawa S, Starikovskaia S, Tennyson J, Terashima K, Turner M~M, van~de Sanden M~C~M and Vardelle A 2017 {\em Journal of Physics D: Applied Physics\/} {\bf 50} 323001 \urlprefix\url{https://dx.doi.org/10.1088/1361-6463/aa76f5}

\bibitem{alves2018foundations}
Alves L, Bogaerts A, Guerra V and Turner M 2018 {\em Plasma Sources Science and Technology\/} {\bf 27} 023002

\bibitem{oehrlein2018foundations}
Oehrlein G~S and Hamaguchi S 2018 {\em Plasma Sources Science and Technology\/} {\bf 27} 023001

\bibitem{von2014clinical}
Von~Woedtke T, Metelmann H~R and Weltmann K~D 2014 {\em Contributions to Plasma Physics\/} {\bf 54} 104--117

\bibitem{holste2020ion}
Holste K, Dietz P, Scharmann S, Keil K, Henning T, Zsch{\"a}tzsch D, Reitemeyer M, Nausch{\"u}tt B, Kiefer F, Kunze F {\em et~al.\/} 2020 {\em Review of Scientific Instruments\/} {\bf 91}

\bibitem{samukawa20122012}
Samukawa S, Hori M, Rauf S, Tachibana K, Bruggeman P, Kroesen G, Whitehead J~C, Murphy A~B, Gutsol A~F, Starikovskaia S {\em et~al.\/} 2012 {\em Journal of Physics D: Applied Physics\/} {\bf 45} 253001

\bibitem{kaganovich2020physics}
Kaganovich I~D, Smolyakov A, Raitses Y, Ahedo E, Mikellides I~G, Jorns B, Taccogna F, Gueroult R, Tsikata S, Bourdon A {\em et~al.\/} 2020 {\em Physics of Plasmas\/} {\bf 27}

\bibitem{adam2008physics}
Adam J, Boeuf J~P, Dubuit N, Dudeck M, Garrigues L, Gresillon D, Heron A, Hagelaar G, Kulaev V, Lemoine N {\em et~al.\/} 2008 {\em Plasma Physics and Controlled Fusion\/} {\bf 50} 124041

\bibitem{boeuf2017tutorial}
Boeuf J~P 2017 {\em Journal of Applied Physics\/} {\bf 121}

\bibitem{perales2022hybrid}
Perales-D{\'\i}az J, Dom{\'\i}nguez-V{\'a}zquez A, Fajardo P, Ahedo E, Faraji F, Reza M and Andreussi T 2022 {\em Journal of Applied Physics\/} {\bf 131}

\bibitem{shah2023investigation}
Shah M, Chaudhury B and Bandyopadhyay M 2023 {\em Scientific Reports\/} {\bf 13} 20044

\bibitem{kolev2012physics}
Kolev S, Hagelaar G, Fubiani G and Boeuf J~P 2012 {\em Plasma sources science and technology\/} {\bf 21} 025002

\bibitem{schiesko2012magnetic}
Schiesko L, McNeely P, Franzen P, Fantz U, team N {\em et~al.\/} 2012 {\em Plasma Physics and Controlled Fusion\/} {\bf 54} 105002

\bibitem{garrigues2016appropriate}
Garrigues L, Fubiani G and Boeuf J~P 2016 {\em Journal of Applied Physics\/} {\bf 120}

\bibitem{smolyakov2016fluid}
Smolyakov A, Chapurin O, Frias W, Koshkarov O, Romadanov I, Tang T, Umansky M, Raitses Y, Kaganovich I and Lakhin V 2016 {\em Plasma Physics and Controlled Fusion\/} {\bf 59} 014041

\bibitem{hara2019overview}
Hara K 2019 {\em Plasma Sources Science and Technology\/} {\bf 28} 044001

\bibitem{donko2021edupic}
Donk{\'o} Z, Derzsi A, Vass M, Horv{\'a}th B, Wilczek S, Hartmann B and Hartmann P 2021 {\em Plasma Sources Science and Technology\/} {\bf 30} 095017

\bibitem{arber2015contemporary}
Arber T, Bennett K, Brady C, Lawrence-Douglas A, Ramsay M, Sircombe N~J, Gillies P, Evans R, Schmitz H, Bell A {\em et~al.\/} 2015 {\em Plasma Physics and Controlled Fusion\/} {\bf 57} 113001

\bibitem{decyk2014particle}
Decyk V~K and Singh T~V 2014 {\em Computer Physics Communications\/} {\bf 185} 708--719

\bibitem{burau2010picongpu}
Burau H, Widera R, H{\"o}nig W, Juckeland G, Debus A, Kluge T, Schramm U, Cowan T~E, Sauerbrey R and Bussmann M 2010 {\em IEEE Transactions on Plasma Science\/} {\bf 38} 2831--2839

\bibitem{buneman1959dissipation}
Buneman O 1959 {\em Physical Review\/} {\bf 115} 503

\bibitem{6808530}
Langdon A~B 2014 {\em IEEE Transactions on Plasma Science\/} {\bf 42} 1317--1320

\bibitem{derouillat2018smilei}
Derouillat J, Beck A, P{\'e}rez F, Vinci T, Chiaramello M, Grassi A, Fl{\'e} M, Bouchard G, Plotnikov I, Aunai N {\em et~al.\/} 2018 {\em Computer Physics Communications\/} {\bf 222} 351--373

\bibitem{birdsall2004plasma}
Birdsall C~K and Langdon A~B 2004 {\em Plasma physics via computer simulation\/} (CRC press)

\bibitem{SUN201635}
Sun A, Becker M~M and Loffhagen D 2016 {\em Computer Physics Communications\/} {\bf 206} 35--44 ISSN 0010-4655

\bibitem{tskhakaya2007particle}
Tskhakaya D, Matyash K, Schneider R and Taccogna F 2007 {\em Contributions to Plasma Physics\/} {\bf 47} 563--594

\bibitem{chaudhury2019hybrid}
Chaudhury B, Shah M, Parekh U, Gandhi H, Desai P, Shah K, Phadnis A, Shah M, Bandyopadhyay M and Chakraborty A 2019 Hybrid parallelization of particle in cell monte carlo collision (pic-mcc) algorithm for simulation of low temperature plasmas {\em Proceedings Software Challenges to Exascale Computing, SCEC 2018\/} (Springer) pp 32--53

\bibitem{BChaudhury2017}
Shah H, Kamaria S, Markandeya R, Shah M and Chaudhury B 2017 A novel implementation of 2d3v particle-in-cell (pic) algorithm for kepler gpu architecture {\em 2017 IEEE 24th International Conference on High Performance Computing (HiPC)\/} pp 378--387

\bibitem{adams2007performance}
Adams M, Ethier S and Wichmann N 2007 Performance of particle in cell methods on highly concurrent computational architectures {\em Journal of Physics: Conference Series\/} vol~78 (IOP Publishing) p 012001

\bibitem{charoy20192d}
Charoy T, Boeuf J~P, Bourdon A, Carlsson J~A, Chabert P, Cuenot B, Eremin D, Garrigues L, Hara K, Kaganovich I~D {\em et~al.\/} 2019 {\em Plasma Sources Science and Technology\/} {\bf 28} 105010

\bibitem{miller2021dynamic}
Miller K~G, Lee R~P, Tableman A, Helm A, Fonseca R~A, Decyk V~K and Mori W~B 2021 {\em Computer Physics Communications\/} {\bf 259} 107633

\bibitem{deluzet2023efficient}
Deluzet F, Fubiani G, Garrigues L, Guillet C and Narski J 2023 {\em Journal of Computational Physics\/} {\bf 480} 112022

\bibitem{tang2016extreme}
Tang W, Wang B, Ethier S, Kwasniewski G, Hoefler T, Ibrahim K~Z, Madduri K, Williams S, Oliker L, Rosales-Fernandez C {\em et~al.\/} 2016 Extreme scale plasma turbulence simulations on top supercomputers worldwide {\em SC'16: Proceedings of the International Conference for High Performance Computing, Networking, Storage and Analysis\/} (IEEE) pp 502--513

\bibitem{wright2024developing}
Wright S~A, Ridgers C~P, Mudalige G~R, Lantra Z, Williams J, Sunderland A, Thorne H~S and Arter W 2024 {\em Computer Physics Communications\/} {\bf 298} 109123

\bibitem{gruber2023llama}
Gruber B~M, Amadio G, Blomer J, Matthes A, Widera R and Bussmann M 2023 {\em Software: Practice and Experience\/} {\bf 53} 115--141

\bibitem{incardona2019openfpm}
Incardona P, Leo A, Zaluzhnyi Y, Ramaswamy R and Sbalzarini I~F 2019 {\em Computer Physics Communications\/} {\bf 241} 155--177

\bibitem{Hur_2019}
Hur M~Y, Kim J~S, Song I~C, Verboncoeur J~P and Lee H~J 2019 {\em Plasma Research Express\/} {\bf 1} 015016 \urlprefix\url{https://dx.doi.org/10.1088/2516-1067/ab0918}

\bibitem{taccogna2023plasma}
Taccogna F, Cichocki F, Eremin D, Fubiani G and Garrigues L 2023 {\em Journal of Applied Physics\/} {\bf 134}

\bibitem{jambunathan2018chaos}
Jambunathan R and Levin D~A 2018 {\em Journal of computational physics\/} {\bf 373} 571--604

\bibitem{juhasz2021efficient}
Juhasz Z, {\v{D}}urian J, Derzsi A, Matej{\v{c}}{\'\i}k {\v{S}}, Donk{\'o} Z and Hartmann P 2021 {\em Computer Physics Communications\/} {\bf 263} 107913

\bibitem{LIEWER1989302}
Liewer P~C and Decyk V~K 1989 {\em Journal of Computational Physics\/} {\bf 85} 302--322 ISSN 0021-9991

\bibitem{vincenti2017efficient}
Vincenti H, Lobet M, Lehe R, Sasanka R and Vay J~L 2017 {\em Computer Physics Communications\/} {\bf 210} 145--154

\bibitem{steiniger2023ez}
Steiniger K, Widera R, Bastrakov S, Bussmann M, Chandrasekaran S, Hernandez B, Holsapple K, Huebl A, Juckeland G, Kelling J {\em et~al.\/} 2023 {\em Computer Physics Communications\/} {\bf 291} 108849

\bibitem{claustre2013particle}
Claustre J, Chaudhury B, Fubiani G, Paulin M and Boeuf J~P 2013 {\em IEEE Transactions on plasma science\/} {\bf 41} 391--399

\bibitem{decyk2011adaptable}
Decyk V~K and Singh T~V 2011 {\em Computer Physics Communications\/} {\bf 182} 641--648

\bibitem{abreu2010pic}
Abreu P, Fonseca R~A, Pereira J~M and Silva L~O 2010 {\em IEEE Transactions on Plasma Science\/} {\bf 39} 675--685

\bibitem{Stantchev2008}
Stantchev G, Dorland W and Gumerov N 2008 {\em Journal of Parallel and Distributed Computing\/} {\bf 68}(10) 1339--1349 ISSN 07437315

\bibitem{hockney2021computer}
Hockney R~W and Eastwood J~W 2021 {\em Computer simulation using particles\/} (crc Press)

\bibitem{madduri2012optimization}
Madduri K, Su J, Williams S, Oliker L, Ethier S and Yelick K 2012 {\em IEEE Transactions on Parallel and Distributed Systems\/} {\bf 23} 1915--1922

\bibitem{peskin1989three}
Peskin C~S and McQueen D~M 1989 {\em Journal of Computational Physics\/} {\bf 81} 372--405

\bibitem{sorensen2008accelerating}
S{\o}rensen T~S, Schaeffter T, Noe K~{\O} and Hansen M~S 2008 {\em IEEE Transactions on Medical Imaging\/} {\bf 27} 538--547

\bibitem{hara2023effects}
Hara K, Robertson T, Kenney J and Rauf S 2023 {\em Plasma Sources Science and Technology\/} {\bf 32} 015008

\bibitem{donko2011particle}
Donk{\'o} Z 2011 {\em Plasma Sources Science and Technology\/} {\bf 20} 024001

\bibitem{schenk2004solving}
Schenk O and G{\"a}rtner K 2004 {\em Future Generation Computer Systems\/} {\bf 20} 475--487

\bibitem{boris1970relativistic}
Boris J~P {\em et~al.\/} 1970 Relativistic plasma simulation-optimization of a hybrid code {\em Proc. Fourth Conf. Num. Sim. Plasmas\/} pp 3--67

\bibitem{turner2013simulation}
Turner M~M, Derzsi A, Donko Z, Eremin D, Kelly S~J, Lafleur T and Mussenbrock T 2013 {\em Physics of Plasmas\/} {\bf 20}

\bibitem{villafana20212d}
Villafana W, Petronio F, Denig A, Jimenez M, Eremin D, Garrigues L, Taccogna F, Alvarez-Laguna A, Boeuf J~P, Bourdon A {\em et~al.\/} 2021 {\em Plasma Sources Science and Technology\/} {\bf 30} 075002

\bibitem{alves2023foundations}
Alves L~L, Becker M~M, van Dijk J, Gans T, Go D~B, Stapelmann K, Tennyson J, Turner M~M and Kushner M~J 2023 {\em Plasma Sources Science and Technology\/} {\bf 32} 023001

\bibitem{boeuf2018b}
Boeuf J~P and Garrigues L 2018 {\em Physics of Plasmas\/} {\bf 25}

\bibitem{villafana20233d}
Villafana W, Cuenot B and Vermorel O 2023 {\em Physics of Plasmas\/} {\bf 30}

\bibitem{charoy2020comparison}
Charoy T, Lafleur T, Tavant A, Chabert P and Bourdon A 2020 {\em Physics of Plasmas\/} {\bf 27}

\bibitem{reza2023concept}
Reza M, Faraji F and Knoll A 2023 {\em Journal of Physics D: Applied Physics\/} {\bf 56} 175201

\bibitem{qiang2010particle}
Qiang J and Li X 2010 {\em Computer Physics Communications\/} {\bf 181} 2024--2034

\bibitem{wang2019modern}
Wang B, Ethier S, Tang W, Ibrahim K~Z, Madduri K, Williams S and Oliker L 2019 {\em The International Journal of High Performance Computing Applications\/} {\bf 33} 169--188

\bibitem{bird2021vpic}
Bird R, Tan N, Luedtke S~V, Harrell S~L, Taufer M and Albright B 2021 {\em IEEE Transactions on Parallel and Distributed Systems\/} {\bf 33} 952--963

\bibitem{hara2020cross}
Hara K and Tsikata S 2020 {\em Physical Review E\/} {\bf 102} 023202

\bibitem{kumar2021effects}
Kumar P, Tsikata S and Hara K 2021 {\em Journal of Applied Physics\/} {\bf 130}

\bibitem{birdsall1991particle}
Birdsall C~K 1991 {\em IEEE Transactions on plasma science\/} {\bf 19} 65--85

\bibitem{MERTMANN20112161}
Mertmann P, Eremin D, Mussenbrock T, Brinkmann R~P and Awakowicz P 2011 {\em Computer Physics Communications\/} {\bf 182} 2161--2167 ISSN 0010-4655

\bibitem{KHAZIEV201887}
Khaziev R and Curreli D 2018 {\em Computer Physics Communications\/} {\bf 229} 87--98 ISSN 0010-4655

\end{thebibliography}

\end{document}